*Original Article*

# The Influence of Tourist Experience on Revisit Decisions with the Mediation of Tourist Satisfaction

[1]**Marsuni H. Muhammad**, [2]**Ida Hidayanti**, [3]**Sulfi Abdul Haji**, [4]**Rahmat Sabuhari**
[1,2,3,4]*Management, Faculty of Economic and Business, Universitas Khairun, Ternate, Indonesia.*

***Abstract:*** *Nusliko Park Ecotourism Area is a combination of natural and artificial tourism with the background of Lake Nusliko, which can be developed as a leading tourist attraction. Marketing strategies can be formulated by analyzing the factors influencing the decision to revisit. The decision to revisit is a key indicator that shows the success of a tourism marketing strategy. This study aimed to test and analyze the relationship between tourist experience and tourist satisfaction with the decision to revisit the Nusliko Park Ecotourism area. This type of research is descriptive research. The population in this study was visitors to the Nusliko Park Ecotourism area. The sample criteria were visitors who had visited the Nusliko Park Ecotourism area at least twice and were domiciled in North Maluku. Data were collected through questionnaires, and 114 questionnaires were returned. The research method used non-probability sampling with purposive sampling techniques. These two techniques were used to make it easier for the author to obtain samples because the Nusliko Park Ecotourism area was closed when the research was conducted. After all, it was under repair. Data were processed and analyzed using SEM and through SmartPLS 4.0 software. The study results indicate that the tourist experience and satisfaction positively influence the decision to revisit. Through this study, the satisfaction variable was also found to mediate the relationship between the tourist experience and the decision to revisit.*

***Keywords:*** *Tourist experience, Tourist satisfaction, Revisit decision, Ecotourism area, Nusliko Park.*

## I. INTRODUCTION

Tourism is a multidimensional sector that involves various actors across sectors and demands a holistic approach to its management. This sector has driven economic growth in many developing countries by contributing to national income, job creation, and poverty alleviation. However, the development of the tourism industry is often accompanied by negative impacts on the environment and local socio-cultural values. Therefore, sustainable tourism that balances economic, social, and environmental dimensions has become the leading paradigm in developing contemporary tourist destinations (Utomo & Afran, 2023).

In this context, ecotourism emerges as a form of sustainable tourism that emphasizes environmental conservation, empowerment of local communities, and educational experiences for tourists. Ecotourism is not only considered a solution to environmental degradation due to mass tourism but also a means of economic diversification that supports the welfare of local communities (Aulia, 2023). Susilawati (2016) stated that ecotourism products should integrate educational, socio-cultural, aesthetic, ethical, and reputational dimensions so that ecotourism planning must be based on strong conservation principles and a mature understanding of the market.

One of the destinations built with this approach is the Nusliko Park Ecotourism Area, which is strategically located in Central Halmahera Regency, North Maluku. This area is designed for environmental conservation, especially the mangrove ecosystem, and to encourage local economic growth through job creation, business opportunities, and increasing regional income. Since its inauguration in 2020, this area has shown significant dynamics in tourist visits, with the highest spike occurring in 2021, before experiencing fluctuations due to the COVID-19 pandemic and infrastructure improvements.

Although the physical facilities in this area are relatively adequate—such as boardwalks, observation towers, floating cottages, and various tourist attractions—the management of the area still faces several challenges. The main problems identified are weak governance and suboptimal marketing strategies, mainly due to limited human resources in the tourism sector. As a result, the level of repeat tourist visits is still unstable, and the area's potential as a leading destination has not been fully realized.

The propensity of tourists to return (revisit intention) serves as a pivotal metric for assessing the sustainability of a destination and is significantly shaped by two primary factors: the tourism experience and the level of tourist satisfaction. Within the realm of tourism scholarship, Pine and Gilmore's (1998) theoretical framework concerning the experience economy posits that experiences that encompass elements of entertainment, aesthetics, education, and escapism are likely to forge profound impressions that subsequently foster tourist loyalty. In relation to Nusliko Park, these four dimensions are inherently integrated and have the potential to serve as a competitive advantage for the destination.

Moreover, tourist satisfaction has been acknowledged as a significant mediating variable that enhances the correlation between experience and the intention to revisit. Studies by Severt et al. (2007) and Chen & Chen (2010) emphasize that satisfaction arises from comparing tourists' initial expectations and actual experiences during the visit. If expectations are met or exceeded, satisfaction increases, and vice versa. Therefore, a deep understanding of tourist experience, satisfaction, and its implications for revisit intention is crucial for sustainable destination management.

This research undertakes a case study within the Nusliko Park Ecotourism region to examine the impact of tourist experiences and satisfaction on the propensity for revisiting. In particular, this research additionally investigates the mediating function of satisfaction in enhancing the effect of experiences on tourist loyalty. Given the increasing competition between destinations in Central Halmahera Regency, the findings of this study are expected to provide evidence-based strategic recommendations for local governments and other tourism stakeholders to improve destinations' competitiveness and sustainability in the long term.

## II. LITERATURE REVIEW

### A) *The Relationship between Tourist Experience and the Decision to Revisit Ecotourism Areas*

This study is based on two main theoretical frameworks: Experiential Theory (Pine & Gilmore, 1998). This theory states that the tourist experience has four dimensions: Entertainment, Education, Escapism, and Aesthetics. Destinations that activate these four dimensions provide profound and memorable experiences, which can influence tourist behavior and decisions. Second, Expectation-Confirmation Theory (Oliver, 1980). This theory explains that satisfaction is formed when the results of the experience exceed or match the expectations held before the trip. Conversely, if the results do not match expectations, dissatisfaction will arise. High satisfaction can increase the likelihood of tourists making repeat visits.

Tourist experiences have long been recognized as an important determinant of post-visit behavior, especially in ecotourism, which emphasizes authentic interactions with nature and local culture. The Planned Behavior Theory (Ajzen, 1991) states that attitudes, subjective norms, and perceived behavioral control influence behavioral intentions, including revisit intentions. In tourism, positive experiences can shape attitudes that support revisit intentions. In addition, the holistic experience theory emphasizes that experiences that involve emotional, cognitive, and sensory aspects can increase tourist satisfaction and loyalty. Tourism experience is one of the most important elements in the tourism industry.

Empirical research supports a positive relationship between tourist experiences and revisit intentions. Gustari et al. (2022) found that tourist experiences significantly influenced revisit intentions, with memory as an important mediator. Similarly, a study by Dewi and Musmini (2023) showed that tourist experience and satisfaction had a significant relationship with revisit intentions to tourist villages. Other research in Baluran National Park also indicated that tourist experiences influenced satisfaction and revisit intentions.

In the context of ecotourism, immersive and authentic experiences are key to shaping revisit intentions. A study by Leonita et al. (2022) in Ngurah Rai Forest Park showed that motivation and risk perception influence revisit intentions, highlighting the importance of safe and satisfying experiences. Oh et al. (2007) strengthened the validity of this experience model in the context of tourism. Another study by Ali et al. (2016) also showed that the stronger tourists perceive these dimensions, the more likely they feel satisfied and want to return. Overall, the literature suggests that positive and meaningful tourist experiences contribute significantly to the decision to revisit an ecotourism destination.

H1: Tourism experience has a positive and significant influence on tourists' decision to revisit.

### B) *Relationship Between Tourism Experience And Tourist Satisfaction*

In the context of ecotourism, tourist experience is a key element that influences their satisfaction with the destination they visit. This experience includes interactions with nature, local culture, and services provided by destination managers. According to the theory of tourist experience, the quality of these interactions can shape positive perceptions that lead to tourist satisfaction. This satisfaction reflects the fulfillment of expectations and indicates the success of ecotourism destination management.

The investigation conducted by Dewi et al. (2023) within the context of the Penglipuran Tourism Village illustrates that memorable tourism experiences exert a substantial impact on tourist satisfaction. This research underscores the critical necessity of cultivating immersive and authentic experiences in order to enhance visitor satisfaction. Similarly, research by Ramadhani et al. (2022) in Umbul Ponggok, Klaten, found that positive emotional experiences contribute significantly to tourist satisfaction, which in turn can increase their loyalty to the destination.

Furthermore, a study by Ikwanadi (2020) in Ekowisata Kampung Blekok Situbondo revealed that customer experience and servicescape significantly influence consumer satisfaction. This study emphasizes that pleasant experiences and supportive environments can increase tourist satisfaction, encouraging word of mouth. These findings are consistent with the customer satisfaction theory, which states that positive experiences during a tourist. These results align with customer satisfaction theory, which posits that favorable experiences encountered during a tourist visit are pivotal in influencing satisfaction and shaping future behavioral intentions.

H2: Tourism experience has a positive and significant influence on tourist satisfaction.

*C) Relationship between tourist satisfaction and the decision to revisit*

The decision to revisit or revisit intention indicates a destination's success in creating tourist loyalty. According to Chen and Tsai (2007), revisit intention is influenced by previous experiences, satisfaction, and perceived value. A study by Prayag et al. (2017) shows that destinations with authentic experiences and high satisfaction are more likely to be revisited. This is very relevant in the context of community-based ecotourism development. Tourist satisfaction is a crucial factor in determining the success of an ecotourism destination. In ecotourism, satisfaction reflects the fulfillment of tourists' expectations regarding facilities and services and includes emotional experiences and interactions with the natural environment and local culture. Customer satisfaction theory states that high satisfaction levels can encourage positive behavior, including the intention to revisit the same destination. Research by Ristya Primadi and colleagues (2021) at the Grafika Cikole Tourism Terminal shows that consumer satisfaction positively and significantly affects revisit intention. This study emphasizes that the higher the level of visitor satisfaction, the higher the intention to revisit. In addition, research by Herliani Izhar and Yurni Suasti (2023) at the Nagari Tuo Pariangan tourist attraction found that the quality of tourist attractions, quality of service/service, emotions, and convenience simultaneously influenced tourists' intention to revisit. High tourist satisfaction with these aspects encourages them to consider revisiting.

H3: Tourist satisfaction exerts a significant and positive impact on tourists' decision to return to visit.

*D) Tourist satisfaction mediates the tourism experience with tourists' revisit decisions.*

In the context of ecotourism, tourist experiences include interactions with nature, local culture, and services provided by destination managers. Positive experiences can form strong and memorable perceptions, affecting tourist satisfaction. This satisfaction reflects the fulfillment of expectations and indicates the success of ecotourism destination management.

Research by Irsyadi and Andriani (2023) in Sembilan Beach, Sumenep, shows that tourist experiences positively influence the intention to revisit through the mediation of customer satisfaction. The results of this study confirm that tourist satisfaction acts as a mediator in the relationship between tourism experience and intention to revisit. Furthermore, research by Maulana Noto Prasojo (2022) in the Blekok Village Ecotourism, Situbondo, revealed that customer experience and servicescape significantly affect consumer satisfaction and intention to revisit. This study emphasizes that pleasant experiences and a supportive environment can increase tourist satisfaction, encouraging word of mouth.

Based on these findings, tourist satisfaction mediates the influence of tourism experience on the decision to revisit ecotourism areas. Thus, ecotourism destination managers need to focus on improving the quality of tourist experiences to increase satisfaction and encourage repeat visits.

H4: Tourist satisfaction mediates the influence of tourism experience on tourists' decision to revisit in a positive and significant manner.

## III. METHODS

This study uses an explanatory quantitative approach to test the causal relationship between variables, namely tourism experience, tourist satisfaction, and the decision to revisit community-based ecotourism destinations.

The population in this study were tourists who had visited the Nusliko Park ecotourism area of Central Halmahera. The sampling technique used was purposive sampling, with the criteria of respondents who had completed the tour and were willing to complete the questionnaire. The minimum sample size was determined based on the formula of Hair et al. (2010), which is a minimum of 5–10 times the number of indicators. The minimum sample size of 27 indicators in the instrument is 100–200 respondents. The target set is 114 respondents.

Data were collected through a closed questionnaire based on a Likert scale of 1–5, ranging from "Strongly Disagree" to "Strongly Agree." The questionnaire was compiled based on indicators from previous studies (Oh et al., 2007; Ali et al., 2016).

## IV. RESULTS AND DISCUSSION

Convergent validity assesses the extent to which reflective indicators accurately represent their underlying constructs, typically evaluated through the outer loadings of each indicator. The indicator value is valid if the indicator explains the construct variable with a value > 0.7, commonly called the rule of thumb (Ghozali & Latan, 2015). While values below 0.7 should be deleted from the indicator. Convergent validity is established when two independent instruments targeting the same latent construct produce strongly correlated outcomes. Table 1 presents the results of the outer loading analysis.

**Table 1. Outer Loadings**

|  | Tourist Satisfaction | Decision to Revisit | Tourist Experience |
|---|---|---|---|
| X.1 |  |  | 0.838 |
| X.2 |  |  | 0.871 |
| X.3 |  |  | 0.880 |
| X.4 |  |  | 0.843 |
| X.5 |  |  | 0.906 |
| X.6 |  |  | 0.886 |
| X.7 |  |  | 0.883 |
| X.8 |  |  | 0.906 |
| X.9 |  |  | 0.846 |
| X.10 |  |  | 0.798 |
| X.11 |  |  | -0.026 |
| Y.1 |  | 0.926 |  |
| Y.2 |  | 0.898 |  |
| Y.3 |  | 0.929 |  |
| Y.4 |  | 0.924 |  |
| Y.5 |  | 0.910 |  |
| Y.6 |  | 0.929 |  |
| Y.7 |  | 0.849 |  |
| Y.8 |  | 0.955 |  |
| Y.9 |  | 0.921 |  |
| Y.10 |  | 0.921 |  |
| Z.1 | 0.902 |  |  |
| Z.2 | 0.935 |  |  |
| Z.3 | 0.943 |  |  |
| Z.4 | 0.887 |  |  |
| Z.5 | 0.924 |  |  |
| Z.6 | 0.938 |  |  |

***Source:*** *Output SmartPLS 4.0, (2025)*

Table 1. Validity testing shows that all indicators/items used to measure the variables of tourist experience, tourist satisfaction, and revisit decision have a high level of validity with a value above 0.7. However, there is a tourist experience variable in indicator X.11, which is still less than 0.7. So, it is necessary to delete the indicator and retest it. The second validity test shows that all indicators/items used to measure have a high level of validity with a value above 0.7. This shows that the indicators used can effectively and consistently measure the intended construct.

**Table 2. Reliability and Discriminant Validity**

|  | Cronbach's alpha | Composite reliability | AVE |
|---|---|---|---|
| Tourist Satisfaction | 0,965 | 0,971 | 0,850 |
| Decision to Revisit | 0,979 | 0,981 | 0,840 |
| Tourist Experience | 0,963 | 0,968 | 0,751 |

***Source:*** *Output SmartPLS 4.0, (2025)*

Referring to Table 2, the AVE values exceed the threshold of 0.5, indicating that all constructs—namely tourist experience, tourist satisfaction, and revisit intention—demonstrate convergent validity. This suggests that each construct sufficiently accounts for the variance in its associated indicators. Construct reliability is assessed to determine whether the selected indicators appropriately represent their underlying latent variables. According to Ghozali and Latan (2015, p. 75), composite reliability values exceeding 0.7 are recommended for confirmatory research, while values between 0.6 and 0.7 remain acceptable for exploratory studies. Based on the SmartPLS 4.0 output, all constructs exhibit composite reliability and Cronbach's alpha values above 0.6. Therefore, the measurement model satisfies reliability criteria, aligning with Hair et al. (2014), who suggest that composite reliability or Cronbach's alpha should ideally be above 0.7, though a minimum of 0.6 is acceptable in less stringent contexts.

**Table 3.** *R-Square* **and** *Adjusted R-Square*

|  | R-square | R-square adjusted |
|---|---|---|
| Tourist Satisfaction | 0,635 | 0,632 |
| Decision to Revisit | 0,800 | 0,796 |

*Source: Output SmartPLS 4.0, (2025)*

As presented in Table 3, the R-squared value for the revisit decision is 0.80, indicating that 80% of the variance in revisit intentions is explained by the tourist experience variable, while the remaining 20% is attributable to other factors not included in the current model. Similarly, the R-squared value for tourist satisfaction is 0.63, suggesting that tourist experience accounts for 63% of the variation in satisfaction levels, with the remaining 37% influenced by unexamined variables. The structural model analysis aims to elucidate the relationships among latent variables in the study. The predictive relevance and significance of these relationships are evaluated through t-statistics, which reflect the strength of the associations between independent and dependent variables. The corresponding path coefficients, as obtained from the SmartPLS output, are summarized in Table 4 below.

**Table 4.** *Path Coefficients, Mean, T-Statistic, and P-Value)*

|  | Path Coefficients (O) | Sample mean (M) | T *statistics* | P *values* |
|---|---|---|---|---|
| Tourist Satisfaction -> Decision to Revisit | 0,459 | 0,478 | 2,779 | 0,005 |
| Tourist Experience -> Tourist Satisfaction | 0,797 | 0,798 | 14,772 | 0,000 |
| Tourist Experience -> Decision to Revisit | 0,485 | 0,466 | 2,867 | 0,004 |

*Source: Output SmartPLS 4.0, 2025.*

The path coefficients, also referred to as inner model estimates, represent the strength and significance of the hypothesized relationships between latent variables. The following section presents a detailed interpretation of the path coefficient results:

Table 4. The path coefficient value of 0.485 is obtained; this means that the direction of the relationship between the tourist experience and the decision to revisit is positive. In addition, the T-statistic for hypothesis one, namely the influence of tourist experience on the decision to revisit, is 2.867, with a p-value of 0.004. Thus, the first hypothesis (H1) of this study is accepted. This means that the tourist experience positively influences the decision to revisit. The positive influence between tourist experience and the decision to revisit indicates that the better the tourist experience in the Nusliko Park Ecotourism area, the more it will be able to increase the decision to revisit the tourist attraction and vice versa. This is due to the emotional memory of tourists in the Nusliko Park Ecotourism area and the positive memories generated from previous tourist experiences.

The test results in Table 4. obtained a path coefficient value of 0.797, which means that the direction of the relationship between tourist experience and tourist satisfaction is positive. In addition, the T-statistic for hypothesis two, namely the influence of tourist experience on tourist satisfaction, is 14.772, with a p-value of 0.000. Thus, hypothesis two (H2) in this study is accepted, where tourist experience positively influences tourist satisfaction. The positive influence between tourist experience and satisfaction indicates that the better the experience tourists get when visiting the Nusliko Park Ecotourism area, the greater their satisfaction with the tourist attraction. These results indicate that experience is a key factor in creating tourist satisfaction.

The test results in Table 4. obtained a path coefficient value of 0.459, which means that the direction of the relationship between tourist satisfaction and the decision to revisit is positive. In addition, the T-statistic for hypothesis three, namely the influence of tourist satisfaction on the decision to revisit, is 2.779, with a p-value of 0.005, which is smaller than 0.05 (0.005 <0.05). Thus, hypothesis three (H3) in this study is accepted, where tourist satisfaction positively influences the decision to revisit. The positive influence between tourist satisfaction and the decision to revisit indicates that the more satisfied tourists are with the Nusliko Park Ecotourism area, the greater the tourists will decide to revisit the tourist attraction at a different time.

Mediation analysis is performed to determine the role and position of variables within the model. This testing relies on the significance of parameter estimates presented in Table 5, and mediation involves evaluating both the indirect influence of the independent variable on the dependent variable.

**Table 5. The Role of Tourist Satisfaction as Mediating Variables**

|  | *Path Coefficient* (O) | Sample mean (M) | T *statistics* | P *value* |
|---|---|---|---|---|
| Tourist Experience -> Tourist Satisfaction -> Decision to Revisit | 0,366 | 0,385 | 2,538 | 0,011 |

**Source:** *Output SmartPLS* 4.0, (2025)

The values in Table 5. The following is an explanation of the results of the indirect effect:

The path coefficient value of 0.366 with a T-statistic of 2.538 and a p-value of 0.011 < 0.05 indicates that this relationship is statistically significant. The coefficient value of 0.366 indicates that the influence of tourist experience on the decision to revisit becomes stronger when tourist satisfaction is used as a mediating variable. Thus, hypothesis four (H4) in this study is accepted. The strong influence of tourist satisfaction as a mediating variable between tourist experience and the decision to revisit indicates that positive tourist experiences in the Nusliko Park Ecotourism area not only directly increase tourist satisfaction but also indirectly strengthen tourists' decisions to revisit the tourist attraction by increasing their level of satisfaction first.

The results of the hypothesis testing show that the tourist experience has a positive influence on the decision to revisit. This indicates that the experience tourists have during their visit significantly contributes to their intention to repeat the visit. The sense dimension (five senses), which includes road access, environmental conditions, and price, are the main aspects that shape tourists' initial perceptions.

The feel dimension also significantly contributes, with the most dominant indicator being the feeling of safety and comfort in the tourist environment. The friendliness of the manager, as part of social interaction, also strengthens the positive emotional experience of tourists. Furthermore, the act dimension emphasizes the importance of tourist involvement in activities such as playing water bikes, taking selfies, and exploring mangrove forests as a direct experience that strengthens emotional bonds. The think dimension shows that perceptions of the safety and comfort of the area also provide a positive experience. In contrast, the related dimension has a minor influence, indicating the need to further evaluate the social aspects and the shared values that tourists feel.

This finding aligns with previous studies by Kozak (2001) and Campo et al. (2010), which found that past visit experiences influence the decision to revisit. This study also supports the empirical experience theory of Pine & Gilmore (1998), which emphasizes that experience begins before arrival and ends with memory and future visit plans. In tourism marketing, immersive, safe, and memorable experiences are the primary keys to attracting and retaining tourists. Further analysis shows that the tourist experience also positively affects tourist satisfaction. Indicators such as friendly service, affordable prices, beautiful scenery, and adequate facilities are the main factors in forming a positive experience. These results are consistent with the findings of Wang et al. (2015), Iranita (2017), Pujiastuti et al. (2020), and Prakoso et al. (2020), which confirm a significant influence between experience and tourist satisfaction.

Tourists who feel that their expectations and desires are met are more satisfied. Tourism destination managers must pay attention to details such as environmental cleanliness, staff friendliness, the accuracy of tourism information, and the uniqueness of the attractions. Mossberg (2007) emphasized that experiences are created through active interaction between tourists and destination elements.

A positive experience creates a good perception of the tourist area and strengthens the emotional bond of tourists. In their theory, Severt et al. (2007) explained that satisfaction arises from the interaction between product or service features and tourist expectations. Furthermore, the results of the study showed that tourist satisfaction has a positive influence on the decision to revisit. The main dimensions influencing satisfaction are management services and the suitability of services with rates and expectations. The first interaction of tourists with the manager and the perception of the value for money spent are the main factors in creating an initial impression. Satisfaction derived from a positive experience creates an emotional appeal that encourages the intention to revisit. The suitability of services with expectations is closely related to natural beauty, completeness of facilities, and quality of service, which is based on tourist expectations.

The results of the hypothesis test show that tourist satisfaction plays a significant mediating role in strengthening the relationship between tourism experience and the decision to revisit. These findings support the theoretical framework stating that positive tourism experiences, which include aspects of service, facilities, and destination attractions, will increase the level of satisfaction, which then indirectly influences the decision to revisit. In other words, although positive experiences directly impact the decision to revisit, the influence will be much stronger if accompanied by a deep sense of satisfaction from tourists. This approach is in line with the mediation model in tourism consumer behavior expressed by Oliver (1980), who states that satisfaction is an evaluative form of experience that can shape behavioral intentions in the future.

This finding also strengthens the role of various dimensions of tourism experience, such as strategic location, affordable price, friendly management, and adequate facilities, as the main determinants of satisfaction. The Nusliko Park Ecotourism Area, which is easily accessible and provides facilities such as toilets, prayer rooms, gazebos, and well-maintained water bikes, creates a safe and comfortable atmosphere for visitors. In addition, the friendliness of the management in providing services contributes significantly to the formation of positive experiences, as indicated by the highest mean score on the satisfaction variable related to direct interaction between tourists and destination staff. This finding aligns with the research of Wicaksono and Santoso (2015), which suggests that the availability of adequate physical facilities is crucial for visitor comfort and influences the decision to make repeat visits. Furthermore, the emotional dimension of tourist satisfaction clarifies the relationship between experience quality and revisit intention. Satisfied tourists tend to make repeat visits and form positive perceptions of the destination, strengthening their loyalty. This study confirms that tourist satisfaction is a dominant intermediary variable that bridges the influence of experience on revisit decisions, as emphasized by previous studies such as Kozak (2001), Campo et al. (2010), and Wang et al. (2015). In the context of Nusliko Park, satisfaction is not only formed by functional aspects alone, but also by holistic experiences that encompass the dimensions of "sense, feel, act, and think," as developed in experiential marketing. This finding provides strategic implications for destination management, especially in policy-making and marketing experience-based tourism. Local governments and the Tourism Office need to prioritize creating high-quality experiences through improving infrastructure, adding tourist attractions, and ensuring the safety and comfort of the area. Although the branding aspect can attract initial visits, tourists' decisions to make repeat visits are greatly influenced by the quality of their previous experiences, which create satisfaction. Thus, the Nusliko Park Ecotourism development strategy must create a positive cycle: experience, satisfaction, and loyalty, as recommended in the literature on experience-oriented tourism.

## V. CONCLUSION

This study confirms that the tourist experience plays a significant role in shaping satisfaction and revisiting decisions. Dimensions of experience such as sense, feel, act, and think contribute positively, primarily focusing on the quality of interaction, uniqueness of attractions, and perception of value. Destination managers need to optimize these aspects to improve the competitiveness and sustainability of tourism areas.\

Theoretically, this study strengthens theoretical models that explain the relationship between tourist experience, satisfaction, and behavior, such as empirical experience theory and tourist satisfaction theory. Destination management and marketing strategies should focus on creating meaningful and enjoyable experiences. Nusliko Park Ecotourism area managers should prioritize the tourist experience at the center of their service strategy by considering accessibility, social interaction, memorable attractions, and value creation that aligns with expectations. This will increase satisfaction and motivate repeat visits, which are essential for destination sustainability.